\documentclass{article}
\usepackage{amsmath, amssymb, amsfonts}
\title{Kinematic quantities for a spherical distribution of uniformly accelerated observers}
\author{Hristu Culetu, \\Ovidius University, Dept.of Physics, \\B-dul Mamaia 124, 900527 Constanta, Romania, \\e-mail : hculetu@yahoo.com}

\begin{document}
\numberwithin{equation}{section}
\pagenumbering{arabic}
\maketitle
\newcommand{\fv}{\boldsymbol{f}}
\newcommand{\tv}{\boldsymbol{t}}
\newcommand{\gv}{\boldsymbol{g}}
\newcommand{\OV}{\boldsymbol{O}}
\newcommand{\wv}{\boldsymbol{w}}
\newcommand{\WV}{\boldsymbol{W}}
\newcommand{\NV}{\boldsymbol{N}}
\newcommand{\hv}{\boldsymbol{h}}
\newcommand{\yv}{\boldsymbol{y}}
\newcommand{\RE}{\textrm{Re}}
\newcommand{\IM}{\textrm{Im}}
\newcommand{\rot}{\textrm{rot}}
\newcommand{\dv}{\boldsymbol{d}}
\newcommand{\grad}{\textrm{grad}}
\newcommand{\Tr}{\textrm{Tr}}
\newcommand{\ua}{\uparrow}
\newcommand{\da}{\downarrow}
\newcommand{\ct}{\textrm{const}}
\newcommand{\xv}{\boldsymbol{x}}
\newcommand{\mv}{\boldsymbol{m}}
\newcommand{\rv}{\boldsymbol{r}}
\newcommand{\kv}{\boldsymbol{k}}
\newcommand{\VE}{\boldsymbol{V}}
\newcommand{\sv}{\boldsymbol{s}}
\newcommand{\RV}{\boldsymbol{R}}
\newcommand{\pv}{\boldsymbol{p}}
\newcommand{\PV}{\boldsymbol{P}}
\newcommand{\EV}{\boldsymbol{E}}
\newcommand{\DV}{\boldsymbol{D}}
\newcommand{\BV}{\boldsymbol{B}}
\newcommand{\HV}{\boldsymbol{H}}
\newcommand{\MV}{\boldsymbol{M}}
\newcommand{\be}{\begin{equation}}
\newcommand{\ee}{\end{equation}}
\newcommand{\ba}{\begin{eqnarray}}
\newcommand{\ea}{\end{eqnarray}}
\newcommand{\bq}{\begin{eqnarray*}}
\newcommand{\eq}{\end{eqnarray*}}
\newcommand{\pa}{\partial}
\newcommand{\f}{\frac}
\newcommand{\FV}{\boldsymbol{F}}
\newcommand{\ve}{\boldsymbol{v}}
\newcommand{\AV}{\boldsymbol{A}}
\newcommand{\jv}{\boldsymbol{j}}
\newcommand{\LV}{\boldsymbol{L}}
\newcommand{\SV}{\boldsymbol{S}}
\newcommand{\av}{\boldsymbol{a}}
\newcommand{\qv}{\boldsymbol{q}}
\newcommand{\QV}{\boldsymbol{Q}}
\newcommand{\ev}{\boldsymbol{e}}
\newcommand{\uv}{\boldsymbol{u}}
\newcommand{\KV}{\boldsymbol{K}}
\newcommand{\ro}{\boldsymbol{\rho}}
\newcommand{\si}{\boldsymbol{\sigma}}
\newcommand{\thv}{\boldsymbol{\theta}}
\newcommand{\bv}{\boldsymbol{b}}
\newcommand{\JV}{\boldsymbol{J}}
\newcommand{\nv}{\boldsymbol{n}}
\newcommand{\lv}{\boldsymbol{l}}
\newcommand{\om}{\boldsymbol{\omega}}
\newcommand{\Om}{\boldsymbol{\Omega}}
\newcommand{\Piv}{\boldsymbol{\Pi}}
\newcommand{\UV}{\boldsymbol{U}}
\newcommand{\iv}{\boldsymbol{i}}
\newcommand{\nuv}{\boldsymbol{\nu}}
\newcommand{\muv}{\boldsymbol{\mu}}
\newcommand{\lm}{\boldsymbol{\lambda}}
\newcommand{\Lm}{\boldsymbol{\Lambda}}
\newcommand{\opsi}{\overline{\psi}}
\renewcommand{\tan}{\textrm{tg}}
\renewcommand{\cot}{\textrm{ctg}}
\renewcommand{\sinh}{\textrm{sh}}
\renewcommand{\cosh}{\textrm{ch}}
\renewcommand{\tanh}{\textrm{th}}
\renewcommand{\coth}{\textrm{cth}}

\begin{abstract}
The kinematical quantities derived from the velocity field of a nongeodesic congruence are studied. We found the shear tensor components are finite in time but diverge at the event horizon of the spacetime located at $\rho_{H} = 0$. The surface gravity on the horizon is just the proper acceleration of the uniformly expanding distribution of observers, in spherical Rindler coordinates. The Raychaudhuri equation is fulfilled for our nongeodesic congruence of particles worldlines.
 
\end{abstract}    

\section{Introduction}
 It is well known that the planar Rindler reference frame is endowed with an event horizon of an infinite area. However, defining horizon entropy needs the horizon area be finite. Therefore, to study the thermodynamic properties of Rindler's horizon or the kinematical quantities associated to a congruence of a velocity vector field, we look for other, more convenient coordinates  \cite{HC1} \cite{HS}.
 
 The generalized Rindler reference frame (or the spherical Rindler coordinates) will be used in what follows to address the problem of a nongeodesic congruence of particles with a timelike 4- velocity \cite{ND} \cite{RW}. Even though the Rindler metric is Minkowskian, there will be a distorsion of the particle worldlines, a nonvanishing expansion due to acceleration. In addition, the Raychaudhuri equation \cite{ND} \cite{KSDN} concerning the time evolution of the scalar expansion should be observed.\\
 
 \section{Witten expanding bubble}
  Let us now consider the Witten bubble spacetime \cite{EW} 
  \begin {equation}
  ds^{2} = -g^{2} \rho^{2} dt^{2} + (1-\frac{R^{2}}{\rho^{2}})^{-1} d \rho^{2} + \rho^{2} cosh^{2} gt~ d\Omega^{2} - (1-\frac{R^{2}}{\rho^{2}})~ d \chi^{2} 
 \label {2.1}
 \end{equation}
 with $R~\leq \rho \prec \infty ,~ d \Omega^{2} $ is the metric on the unit 2 - sphere , $R$ and $g$ are constants and $\chi$ - the coordinate of the compactified fifth dimension. Witten has studied the decaying process (an expanding bubble) of the ground state of the Kaluza - Klein geometry, which, although stable classically , is unstable against a semiclassical barrier penetration \cite{SC}. 
 
 Let us take the 4 - dimensional subspace $\chi = const.$ of the Witten bubble geometry \cite{HC3}
 \begin{equation}
   ds^{2} = -g^{2} \rho^{2} dt^{2} + (1-\frac{R^{2}}{\rho^{2}})^{-1} d \rho^{2} + \rho^{2} cosh^{2} gt~ d\Omega^{2} 
 \label{2.2}
 \end{equation}
 The above spacetime is flat provided $\rho >>~R$, but written in spherical Rindler coordinates (a spherical distribution of uniformly accelerated observers uses this type of hyperbolic coordinates) . Let us note that the singularity at $\rho = R$  is a coordinate singularity \cite{BM} as can be seen from the isotropic form of (2.2), using a new radial coordinate $r$ 
 \begin {equation}
 \rho = r + \frac{R^{2}}{4r}
 \label{2.3}
 \end{equation}
 with $R/2 \leq r < \infty$. The metric (2.2) becomes now
 \begin{equation}
 ds^{2} = \left( 1+\frac{R^{2}}{4r^{2}}\right)^{2} (-g^{2} r^{2} dt^{2} + dr^{2} + r^{2} cosh^{2} gt d\Omega^{2}).
 \label{2.4}
 \end{equation} 
  The  Minkowskian character of (2.2) when $\rho >> R$ can be seen from the transformation 
 \begin{equation}
 \bar{x} = \rho cosh gt sin\theta cos \phi ,~~\bar{y} = \rho cosh gt sin \theta sin \phi,~~\bar{z} = \rho cosh gt cos\theta, ~~\bar{t} = \rho sinh gt
 \label{2.5}
 \end{equation}
 where $(\bar{x},\bar{y}, \bar{z}, \bar{t})$ are the Minkowski cartesian coordinates
 \begin{equation}
 ds^{2} = -d \bar{t}^{2} + d \bar{x}^{2} + d \bar{y}^{2} + d \bar{z}^{2} 
 \label{2.6}
 \end{equation}
 and $g$ plays the role of the (constant) acceleration of the spherical distribution of observers.\\
 
 \section{Kinematic parameters} 
 Our purpose in the present paper consists in the determination  of the kinematical quantities (expansion, shear, vorticity, etc.) associated to the 4 - velocity vector field $u^{\alpha}$ in the spacetime
\begin{equation}
   ds^{2} = -g^{2} \rho^{2} dt^{2} + d \rho^{2} + \rho^{2} cosh^{2} gt~ d\Omega^{2} ,
\label{3.1}
\end{equation}
where $u^{\alpha} = (1/g \rho, 0, 0, 0 ),~ u^{\alpha} u_{\alpha} = -1$ . It is interesting to note that the previous ''static'' observers with $\rho = \rho_{0} = const.$ move hyperbolically in Minkowski coordinates 
\begin{equation}
\bar{r}^{2} - \bar{t}^{2} = \rho_{0}^{2} ,
\label{3.2}
\end{equation}
with $\bar{r} = \sqrt{\bar{x}^{2} + \bar{y}^{2} + \bar{z}^{2}}$ , $\bar{r} > \bar{t}$ and the horizon at $\rho = 0$ (as we shall see) corresponds to the light cones $\bar{r} = \pm \bar{t}$. 

 The expansion scalar $\Theta$ of the observers worldlines (the rate of increasing of a volume element) is given by the divergence of $u^{\alpha}$ 
 \begin{equation}
 \Theta \equiv \nabla_{\alpha} u^{\alpha} = \frac{2}{\rho}~ tanh~ gt ,
 \label{3.3}
 \end{equation}
 taken at $\rho = \rho_{0}$. We note that $\Theta$ varies from $-2/\rho$ to $2/\rho$ when $-\infty < t < \infty$. It means the particles' worldlines are expanding for $t > 0$ but undergo a contraction for $t < 0$. The fact that it is an increasing function of time is not surprising if we remind that the spacetime (3.1) is flat, i.e. no gravitation is present to focusing the worldlines. In addition, the congruence is not geodesic and, therefore, the acceleration 4 - vector is nonvanishing. From (3.3) we get
 \begin{equation}
 \dot{\Theta} \equiv u^{\alpha} \nabla_{\alpha} \Theta = \frac{2}{\rho^{2} cosh^{2} gt}
 \label{3.4}
 \end{equation}
 Hence, we have $\dot{\Theta} > 0$ , a consequence of our spherical expansion distribution of uniformly accelerated observers.
 
 Let us compute now the 4 - acceleration of the observers having the velocity $u^{\alpha}$ 
 \begin{equation}
 a^{\alpha} = u^{\beta} \nabla_{\beta} u^{\alpha} 
 \label{3.5}
 \end{equation}
 Using the following nonzero Christoffel symbols
 \begin{equation}
 \Gamma_{tt}^{\rho} = g^{2} \rho,~~\Gamma_{\rho t}^{t} = \frac{1}{\rho}, ~~\Gamma_{\theta \theta}^{t} = \frac{1}{2g} sinh 2gt ,
 \label{3.6}
 \end{equation}
 we obtain $a^{\alpha}  = (0,~1/\rho,~0,~0)$ . The invariant acceleration will be, therefore
 \begin{equation}
 a \equiv  \sqrt{a_{\alpha} a^{\alpha}} = \frac{1}{\rho} .
 \label{3.7}
 \end{equation}
 It is a known fact that the Rindler spacetime has an event horizon \cite{HS} \cite{HC2}. In our coordinates (3.1) the horizon is located at $\rho_{H} = 0$. The surface gravity is, therefore given by
 \begin{equation}
 \kappa = \sqrt{a^{\alpha} a_{\alpha}} \sqrt{-g_{tt}} |_{H} = g .
 \label{3.8}
 \end{equation}
 In other words, the constant $g$ from the coordinate transformation (2.5) is nothing but the surface gravity on the event horizon. It is worth to note that the invariant acceleration $a = 1/\rho$ depends on the position where the ''static'' observer is located. 
 
 Let us find now the shear tensor of the particle worldlines. We have
 \begin{equation}
 \sigma_{\mu \nu} = \frac{1}{2} (h_{\nu}^{\alpha} \nabla_{\alpha} u_{\mu} + h_{\mu}^{\alpha} \nabla_{\alpha} u_{\nu}) - \frac{1}{3} \Theta h_{\mu \nu} + \frac{1}{2} (a_{\mu} u_{\nu} + a_{\nu} u_{\mu}) 
 \label{3.9}
 \end{equation}
 where $h_{\mu \nu} = g_{\mu \nu} + u_{\mu} u_{\nu} $ is the projection tensor onto the direction perpendicular to $u_{\mu}$ and $\sigma_{\mu \nu}$ expresses the distorsion of the worldlines in shape without change in volume \cite{ND}.\\ The eqs. (3.3) and (3.5) yield for the nonzero components
\begin{equation}
\sigma^{\rho}_{ \rho} = -\frac{2}{3 \rho} tanh~ gt, ~~~\sigma_{\theta}^{\theta} = \sigma_{\phi}^{\phi} = \frac{1}{3 \rho} tanh~ gt.
\label{3.10}
\end{equation}
 while 
 \begin{equation}
 \sigma^{2} \equiv \frac{1}{2} \sigma_{\mu \nu} \sigma^{\mu \nu} = \frac{1}{3 \rho^{2}} tanh^{2} gt. 
 \label{3.11}
 \end{equation}

 It can be checked that the shear tensor is traceless. In addition, all the above components  depend on $g$ and vanish when $g$ = 0.

As far as the vorticity tensor 
\begin{equation}
 \omega_{\mu \nu} = \frac{1}{2} (h_{\nu}^{\alpha} \nabla_{\alpha} u_{\mu} - h_{\mu}^{\alpha} \nabla_{\alpha} u_{\nu}) + \frac{1}{2} (a_{\mu} u_{\nu} - a_{\nu} u_{\mu}) 
 \label{3.12}
 \end{equation}
 is concerned, all components are vanishing.
 
 We are now in position to check whether the Raychaudhuri equation 
 \begin{equation}
 \dot{\Theta} - \nabla_{\alpha} a^{\alpha}+ 2(\sigma^{2}- \omega^{2})+ \frac{1}{3} \Theta^{2} = - R_{\alpha \beta} u^{\alpha} u^{\beta}
 \label{3.13}
 \end{equation} 
 is obeyed. Keeping in mind that $\nabla_{\alpha} a^{\alpha} = 2/\rho^{2}, R_{\alpha \beta} =0 $ (our geometry (3.1) is flat) and making use of the previous results, we find that eq.(3.13) holds.
 
 As an overall observation, we note that the expansion scalar, accelaration and some components of the shear tensor are divergent on the horizon. \\
 
 \section{Conclusions}
  We have used a spherical distribution of uniformly accelerated observers to study the kinematic quantities associated to their velocity field vector. The scalar expansion of the worldlines changes sign at $t = 0$ and diverges on the event horizon $\rho = 0$. The fact that $\dot{\Theta}$ is always positive is related to the lack of gravity (our spacetime is Minkowskian in disguise ; it represents the Witten bubble spacetime for $\rho >> R$).
  
  The nongeodesic congruence is endowed with shear whose components vanish when the surface gravity $g$ of the Rindler horizon is null. 
  
  We also checked that the equation for the time evolution of the scalar expansion - the Raychaudhuri equation - is fulfilled.

\end{document}